\documentclass[12pt]{article}
\pdfoutput=1 
\usepackage{graphicx}

\usepackage{booktabs} % For formal tables
\usepackage{enumitem}
\usepackage[noabbrev,capitalize]{cleveref} % For cref
\crefname{equation}{}{} % Custom cref naming
\usepackage{xcolor} % For text coloring
\usepackage{soul} % For highlighting
\usepackage{authblk} % for \affil command

\newcommand{\later}[1]{} % Placeholder for future text
\newcommand{\bet}{{\it beta }} % Italicize 'beta'
\newcommand{\Is}{\mathit{Is}} % Italicize 'Is'

\newlength{\myfigwidth}
\title{Replication-based quantum annealing error mitigation}

\author[$\dagger$,$\star$]{Hristo N.\ Djidjev}
\affil[$\dagger$]{
	Institute of Information and Communication Technologies, Bulgarian Academy of Sciences, Sofia, Bulgaria}
\affil[$\star$]{
	Los Alamos National Laboratory, Los Alamos, NM 87545, USA
}

\begin{document}

\maketitle

\begin{abstract}
	Quantum annealers like those from D-Wave Systems implement adiabatic quantum computing to solve optimization problems, but their analog nature and limited control functionalities present challenges to correcting or mitigating errors. As quantum computing advances towards applications, effective error suppression is an important research goal. We propose a new approach called replication based mitigation (RBM) based on parallel quantum annealing. In RBM, physical qubits representing the same logical qubit are dispersed across different copies of the problem embedded in the hardware. This mitigates hardware biases, is compatible with limited qubit connectivity in current annealers, and is suited for available noisy intermediate-scale quantum (NISQ)  annealers. 
	%Compared to existing repetition-based methods like quantum assisted correction, MREM does not further limit connectivity and is applicable to smaller embeddable problems where solutions are near optimal. It enhances reliability without requiring error-corrected logical qubits.
	Our experimental analysis shows that RBM provides solution quality on par with previous methods while being compatible with a much wider range of hardware connectivity patterns. In comparisons against standard quantum annealing without error mitigation, RBM consistently improves the energies and ground state probabilities across parameterized problem sets. 
	
\end{abstract}

\vspace{1em} % Adds some vertical space for aesthetics

\vspace{1cm}
\section{Introduction}
Quantum computing holds  promise for transforming computation by exploiting the effects of quantum mechanics. Gate based and quantum annealing models have emerged as leading approaches, offering the potential to solve difficult computational problems. However, realizing the full potential of quantum computation requires overcoming errors arising from the fragility of quantum states and the immaturity of the current noisy intermediate-scale quantum (NISQ) technology. Error mitigation represents a significant obstacle faced by all quantum computing platforms. Within this challenge, quantum annealers as those pioneered by D-Wave Systems present a distinctive approach to quantum computing employing adiabatic quantum processes. While having advantages such as much larger number of qubits and relatively easy programming, quantum  annealers also present unique error mitigation challenges due to their analog nature and limited control. 
As quantum computing continues maturing toward practical applications, developing effective strategies for error suppression across diverse technological platforms remains an active area of research.  \textit{Quantum error correction} (QEC) provides essential techniques for protecting quantum information from noise, though approaches differ across paradigms.

%Error correction literature review
A key component of QEC for the gate model are \textit{quantum error correcting codes} (QECCs), which encode each logical qubits as an entangled state of multiple physical qubits \cite{shor1995scheme}. This allows detection and correction of errors on individual physical qubits without destroying the encoded information. %\cite{nielsen2010quantum}. 
\textit{Stabilizer codes} form an important class of QECCs based on stabilizer measurements to identify \textit{error syndromes} \cite{gottesman1997stabilizer}, which are collections of measured outcomes that provide a diagnosis of errors in  QECC without revealing the protected quantum information. %\cite{calderbank1996good}. 
\textit{Topological codes} like surface and color codes are promising for scalability due to high thresholds and efficient syndrome extraction through local operations \cite{kitaev2003fault,bombin2007topological, fowler2012surface}.  

Decoding the syndrome to find a correction is a key challenge in QECC, with efficient decoders needed to maximize practical thresholds \cite{fowler2012towards}. Optimal decoding is computationally intractable, so heuristic decoding algorithms like minimum weight matching provide a practical solution \cite{edmonds1965paths}. Advanced decoders using machine learning and neural networks are also being explored \cite{varsamopoulos2020decoding}. Error suppression techniques like the application of external controllable interactions can reduce physical error rates to below the threshold by mitigating specific noise processes \cite{viola1999dynamical}. Other methods suitable for near-term devices leverage redundancy, specialized measurements, and tailored mappings to detect and correct errors \cite{temme2017error,li2017efficient,endo2018practical,cai2022quantum}.

%Important experimental milestones in QEC include demonstrating repeated QEC cycles \cite{kelly2015state} and fault tolerant operations like state preparation, stabilizer measurement, and magic state injection \cite{chen2021exponential,ryan2022realization,egan2020fault}. Overall, continued progress in QECC development, decoding algorithms, system integration, and fault tolerance is critical for realizing scalable, fault-tolerant quantum computation \cite{fowler2012surface}. Careful benchmarking against realistic noise models will be key for guiding further improvements \cite{ghosh2015noise}.

In quantum annealing error correction, \cite{Pudenz2014} introduces a method called \textit{quantum annealing correction} (QAC) that encodes a single logical qubit by a set of connected physical qubits. At the end of the annealing, the value of the logical qubit is determined by a majority vote. Further developments and analysis of the method were reported in \cite{matsuura2016mean,pearson2019analog}. A generalization called nested \textit{quantum annealing correction} was analyzed in \cite{vinci2016nested, matsuura2019nested}. The \textit{zero-noise extrapolation} method, initially developed for the gate model \cite{endo2018practical,temme2017error}, was applied to quantum annealing in \cite{amin2023quantum}. The idea is that controlled variations in the noise amplitude based on observed responses could allow systematic predictions of the system's behavior under noise-free conditions. A method for utilizing otherwise unused qubits to characterize the current level of noise in quantum annealers was developed in \cite{pelofske2023noise}.

This paper describes a new approach to mitigating errors in quantum annealers called replication based mitigation (RBM) based on the \textit{parallel quantum annealing} (PQA) paradigm \cite{Pelofske2022parallel} and analyzes its performance. The fundamental premise of RBM lies in the simultaneous resolution of multiple instances of a given problem embedded onto the quantum hardware through a single annealing process. The goal is to return the solution with the minimum energy, effectively addressing errors and enhancing the overall reliability of quantum annealing computations.  Compared to the repetition-based approach of QAC 
\cite{Pudenz2014,matsuura2016mean,pearson2019analog,vinci2016nested, matsuura2019nested}, 
where several connected physical qubits represent a single logical one, in RBM the physical qubits corresponding to the logical one are not connected but dispersed to different copies (replicas) of the problem. 
RBM has the following beneficial features. 
(i) It mitigates hardware biases. Quantum hardware inherently introduces biases that can affect the accuracy of solutions. RBM, by solving multiple instances in randomly chosen areas of the quantum chip, allows for the cancellation and mitigation of these biases, leading to more robust and reliable outcomes. This is critical in the current NISQ technology where maintaining uniformity across qubits poses challenges.
(ii) It is better suited to contemporary quantum annealers with limited qubit connectivity compared to alternative methods. Using a set of connected qubits as a single error-corrected logical qubit in QAC and similar methods further lowers the degree of connectivity of the resulting structure and, consequently, the classes of problems that are solvable by the method. 
(iii) While newer generations of quantum annealers boast increasingly larger number of qubits, with the currently newest D-Wave Advantage computers featuring over 5000 qubits, solving the largest embeddable problem often remains infeasible as the hardness of the problems typically increases exponentially with size. 
However, for smaller problems, quantum annealers are more likely to find optimal or near-optimal solutions using the current quantum hardware. 
Hence, the proposed RBM method, being designed for problems smaller than the largest embeddable ones, is well-suited for leveraging the strengths of NISQ-era quantum annealers, which have a steadily increasing number of qubits but still lack sufficient quality for reliably solving the most challenging problems. 
%This allows meaningful gains to be achieved as quantum hardware continues developing on its trajectory toward reduced noise and increased scalability. 

The paper is organized as follows. In Section~\ref{sec:methods}, Methods, we provide background on quantum annealing and describe the proposed RBM method as well as the QAC approach used for comparison. Section~\ref{sec:experiments} presents our experimental analysis, including the setup and benchmarks comparing RBM against QAC and standard QA. Finally, Section~\ref{sec:conclusion} summarizes the key conclusions and discusses directions for further research.

\section{Methods}\label{sec:methods}
\subsection{Quantum annealing}
Quantum annealing (QA) is designed to solve quadratic optimization problems, such as the \textit{Ising problem} represented by:
\begin{equation}
	\mbox{minimize} \; \mathit{Is}(s_1,\dots,s_n) = \sum_{i<j}J_{ij}s_is_j+\sum_ih_is_i,\label{eq:Ising1}
\end{equation}
where $s_i\in\{-1,1\}$ are the variables and $J_{ij}$ and $h_i$ are the problem coefficients. If restricted to $s_i\in\{0,1\}$, it becomes a \textit{quadratic unconstrained binary optimization} (\textit{QUBO}) problem. D-Wave hardware natively implements the Ising version. Many NP-hard problems can be easily reformulated as Ising or QUBO problems \cite{Lucas2014}, which allows them to be solved on the D-Wave devices.

\begin{figure}
	\centering
	\includegraphics[width=0.4\textwidth]{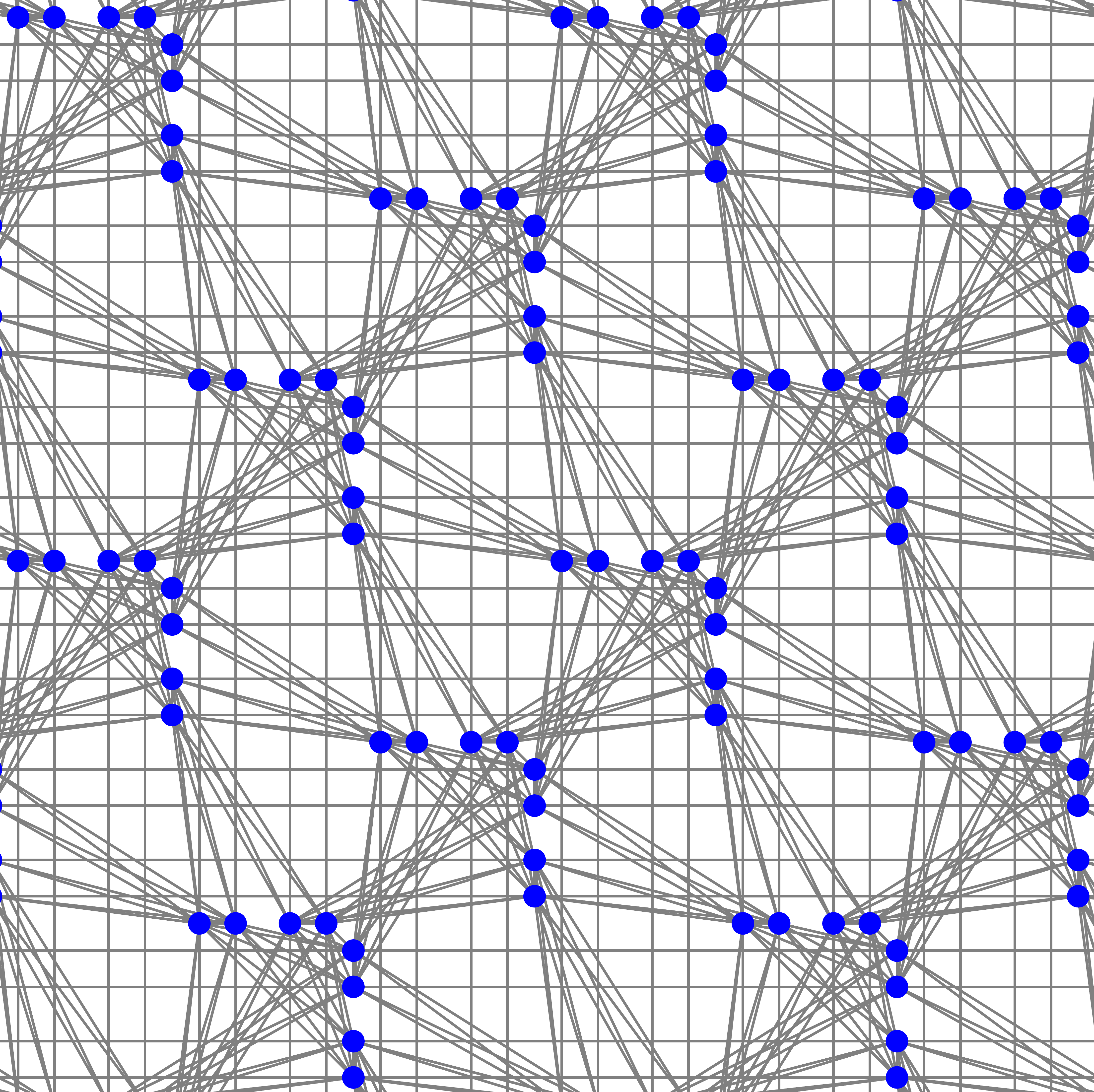}
	\caption{The structure of a Pegasus graph, which defines the connection pattern between  qubits in a D-Wave Advantage quantum chip. Blue dots are nodes (qubits) and connecting edges are couplers.} \label{fig:pegasus_cropped}
\end{figure}

To solve an Ising problem on a D-Wave quantum annealer, coefficients $J_{ij}$ and $h_i$ are sent to the device so that each linear bias $h_i$ is mapped to a distinct qubit $q_i$ and each quadratic bias $J_{ij}$ is mapped to the coupler between qubits $q_i$ and $q_j$.  
However, current D-Wave devices have limited qubit connectivity, see \cref{fig:pegasus_cropped}, which means that most practical problems cannot be natively mapped to the device's quantum processing unit (QPU). In such cases, the graph $I$ describing the structure of the Ising problem, with vertices $\{1,\dots,N\}$ and edges $\{(i,j)~|~J_{ij}\neq 0\}$, which we refer to as the \textit{Ising graph}, is \textit{minor-embedded} into the QPU's \textit{hardware graph} $H$ so that each vertex of $I$ is mapped to a set of connected vertices of $H$, referred as \textit{chains}. (Note that chains are also referred to as logical qubits in the literature, but to avoid confusion, in this paper we use the term logical qubit to refer to an error-corrected qubit and the term chain for the representation of qubits in minor embeddings.) Although chains enable the embedding of Ising problems with arbitrary structures onto the QPU, given a sufficient number of qubits \cite{choi2008minor,choi2011minor}, the presence of long chains introduces additional errors through \textit{freeze-out} \cite{boothby2020next}. This phenomenon occurs when the chains become trapped in a classical state before the completion of the quantum annealing process.

\subsection{The RBM method}
Given an Ising problem $\Is$, we can solve it using the proposed RBM method for error mitigation as follows. (i) We  compute the maximum number of copies, $k$, of $\Is$ that can be embedded onto the QPU based on the problem size and the hardware-graph size and connectivity, and store these $k$ embeddings. 
We discuss below how to find such embeddings in some specific cases.
(ii) We generate a new Ising problem, denoted by $\Is^{(k)}$, containing $k$ identical copies of $\Is$ and embed $\Is^{(k)}$ into the QPU using the stored embeddings.
(iii) We solve $\Is^{(k)}$ by the quantum annealer in a single annealing cycle.
We remind that a a D-Wave annealing call typically returns hundreds or even thousands of samples, as specified by a user-controlled parameter. The rationale is that programming the chip typically takes much more time than a single anneal does, so it is cost-effective to do multiple anneals once the chip is programmed. Let $M$ denote the set of samples (proposed solutions) returned.
(iv) We split each sample of $M$ into $k$ subsamples corresponding each of the $k$ replicas of $\Is$. For each subsample, we compute the value of $\Is$ for the corresponding variable assignment. We use the assignment with the minimum value over all samples $s$ of $M$ and all subsamples of $s$ as a solution for $\Is$.

%journal: include pseudocode
\subsection{The QAC method}
To analyze the effectiveness of RBM, we will compare it against the QAC method \cite{Pudenz2014,pearson2019analog,vinci2015quantum,vinci2016nested}. The QAC method  contains the following elements:
(i) Encoding. The optimization problem is encoded by mapping each logical qubit, $q$, to $k$ physical qubits $ph(q)=\{q_1,\dots,q_k\}$ using repetition codes. 
(ii) Energy penalty for errors. A penalty energy term  is added that raises the energy of states where the $k$ physical qubits in $ph(q)$ take different values. This penalizes errors and suppresses undesirable excitations out of the ground state.
(iii) Decoding. Each logical qubit $q$ is decoded by performing a majority vote on the values the $k$ physical qubits in $ph(q)$ take.

\begin{figure}
	\centering
	\includegraphics[width=0.25\textwidth]{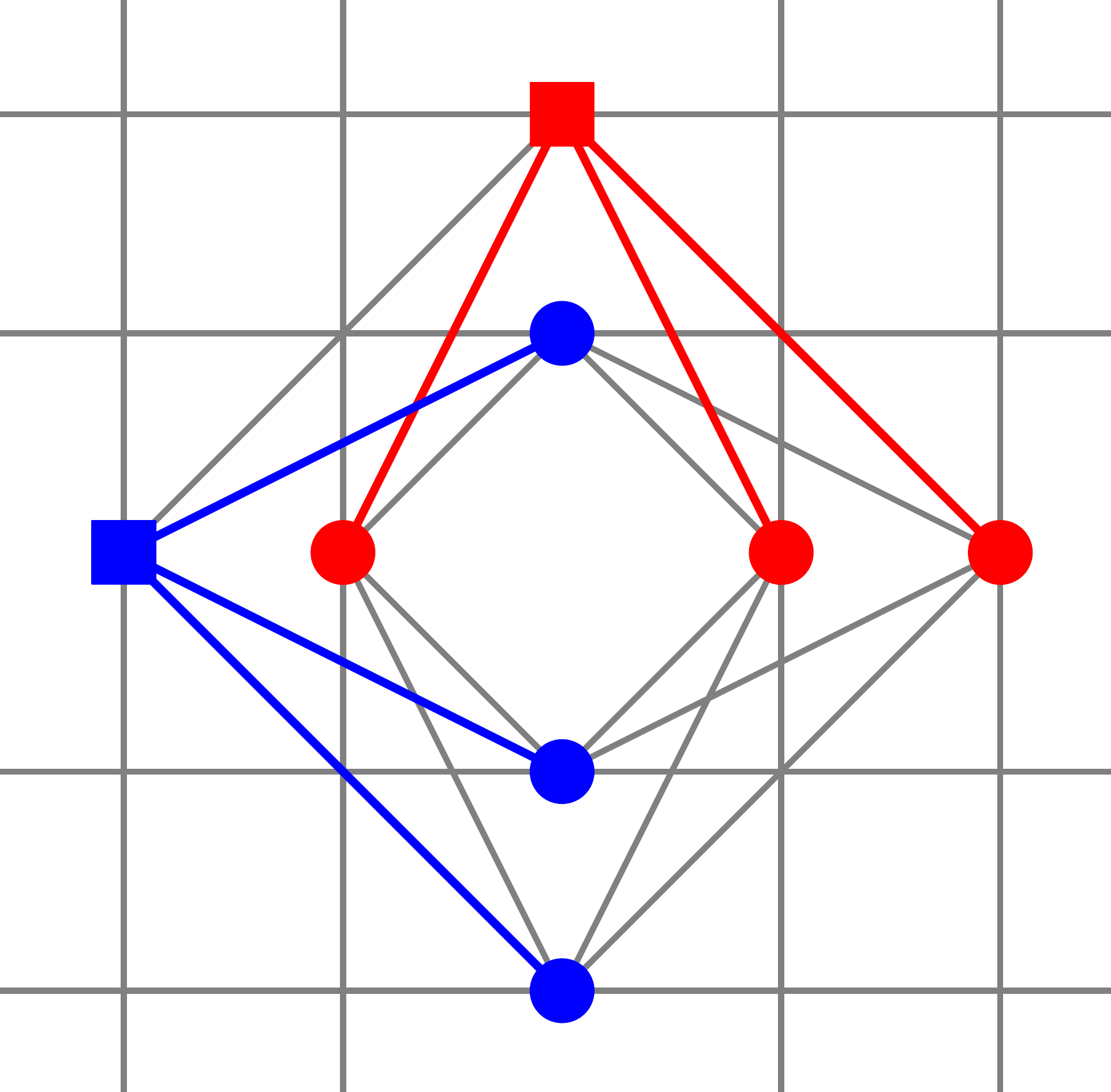}
	\caption{Two QAC logical qubits, shown in different colors, embedded in a cell of the Chimera graph. The penalty qubits are drawn as squares and the problem qubits as circles.}
	\label{fig:chimera_PAC_cropped}
\end{figure}

Specifically, $k=4$ was used in \cite{Pudenz2014,pearson2019analog} for the D-Wave One, 2X, and 2000Q machines whose QPUs use the Chimera graph topology. In this configuration, each logical qubit is represented by a $K_{1,3}$ graph, comprising three \textit{problem qubits} and one \textit{penalty qubit} (\cref{fig:chimera_PAC_cropped}). The penalty qubit exclusively interacts with the problem qubits within the same logical qubit, while the problem qubits also engage with those from other logical qubits. The primary role of the penalty qubit is to encourage its problem qubits to adopt the same value, thereby reducing the effects of single-qubit errors, which are the quantum analogue of bit flips in classical computing. Specifically, in instances where the values differ within a sample, the penalty qubit $p_i$ and one of the problem qubits, say $q_i$, assume different values. 
Since the penalty qubit is connected to its three problem qubits through couplers carrying negative \textit{penalty weight} $\alpha<0$, whose precise value is a tunable parameter,
such discrepancy yields a positive penalty term $p_iq_i\alpha>0$ in the Ising problem, causing its value to deviate from the optimal solution.  

Since any Chimera graph can be embedded in a Pegasus graph of the same order, the embedding from Figure~\ref{fig:chimera_PAC_cropped} can be modified into an embedding into a Pegasus graph, which is the graph used for connectivity in the current D-Wave QPUs. In the next subsection we discuss the embedding aspects in more detail.

\subsection{Generating native embeddings compatible with RBM}
\begin{figure}
	\centering
	\includegraphics[width=0.6\textwidth]{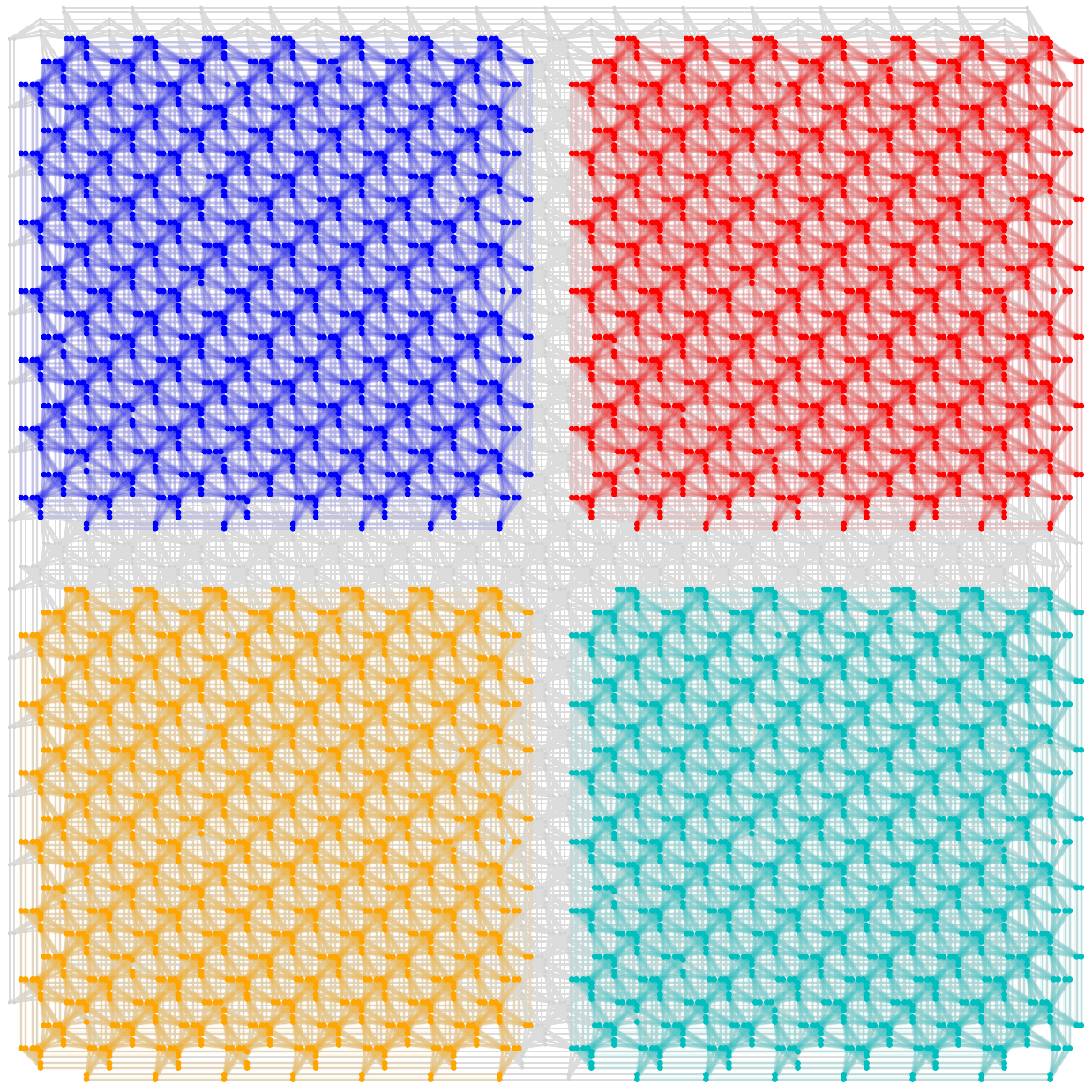}
	\caption{Structure of the Ising graph for the RBM method. Each replica is shown in a different color.
	}
	\label{fig:PQA_noQEC_subgraphs}
\end{figure}

We want to generate \textit{natively} embedded (i.e., without chains) Ising problems that will be used to compare RBM against standard QA without error mitigation. 
For the RBM method, we seek to create $k$ isomorphic disjoint graphs that can be embedded within the QPU hardware graph. In this study, we explore values of $k$ within the set $\{2, 4, 8\}$. Leveraging the symmetrical properties inherent in the Pegasus graph, we can easily partition it into equal parts, maximizing the utilization of available qubits and couplers. 
The resulting Ising graphs for $k=4$ are shown on \cref{fig:PQA_noQEC_subgraphs}. Each graph has $1219$ nodes and $8259$ edges.

However, these Ising problems have structure incompatible with the one required for the QAC method, making these graphs unsuitable for comparing RBM with the QAC method. The next subsection will focus on seeking embeddings that are compatible with both methods.

\subsection{Generating native embeddings compatible with both QAC and RBM}
To assess RBM's performance against QAC and uncorrected QA on identical problem sets, we need problem structures compatible with both error mitigation methods. This entails finding problems that can be embedded in the QPU chip for both RBM and QAC approaches.
%First we consider the case of native embeddings, i.e., when each chain is just a single physical qubit. We will discuss the non-native case in the next subsection.

In order to get a native QAC embedding in the Pegasus graph (\cref{fig:pegasus_cropped}), we should cover it with $K_{1,3}$ graphs, each representing a logical qubit. The  Ising problem will then have a variable and a linear term for each logical qubit and a quadratic term for each pair of logical qubits whose problem physical qubits are connected in the Pegasus topology. On the other hand, in order to have a native RBM embedding, we need to have $k=4$ identical copies of the Ising graph embedded in different regions of the hardware graph. The third constraint we have to take into consideration is that, due to manufacturing defects,  not all vertices and edges of the Pegasus graph correspond to working qubits and couplers in the physical device. Even a single inactive qubit or coupler breaks the symmetry in the Pegasus graph and may result to a large number of additional qubits not being used in our experiment.

\begin{figure}
	\centering
	\includegraphics[width=0.35\textwidth]{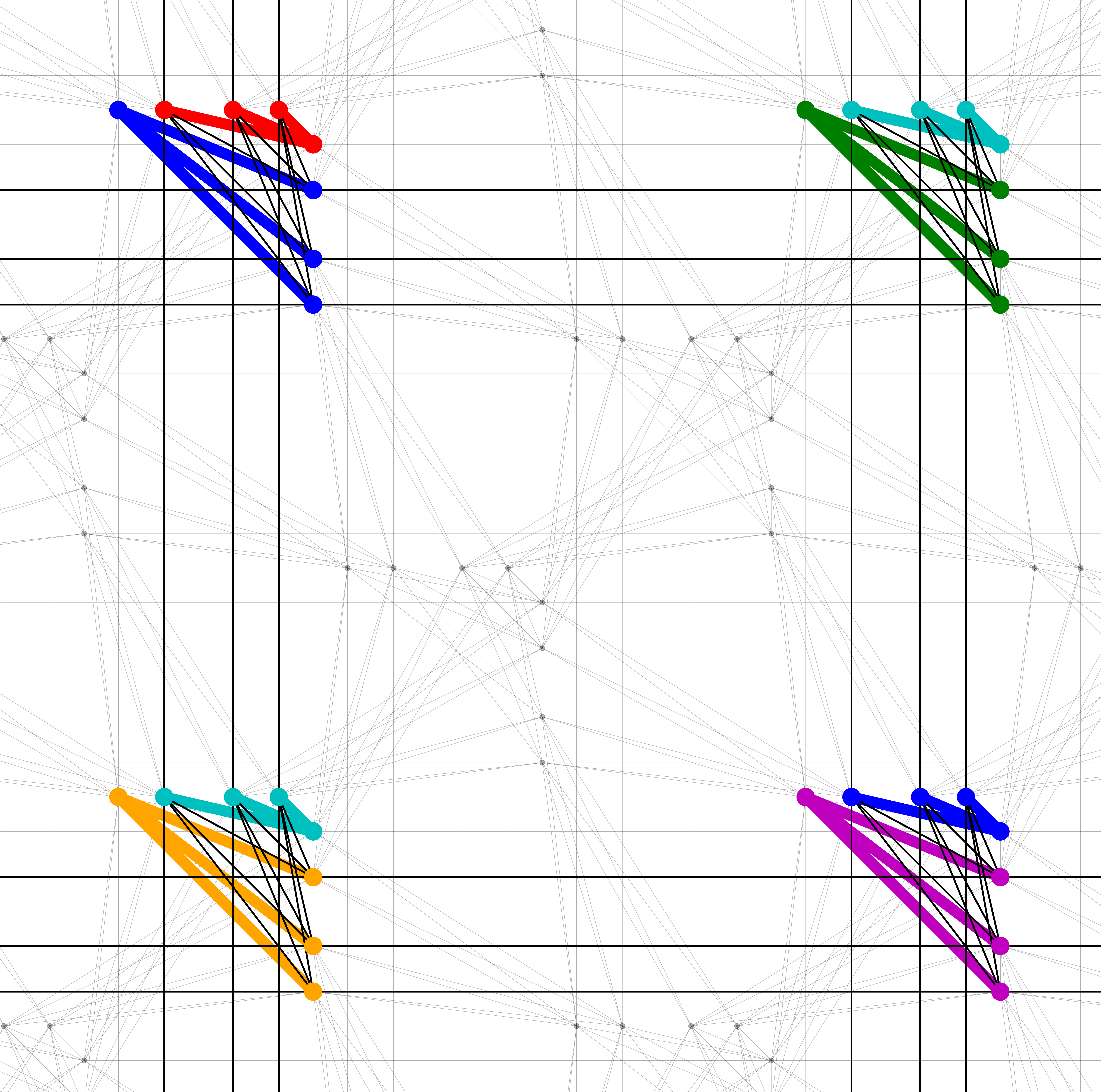}
	\caption{Embedding of the QAC logical qubits in the Pegasus graph. Each logical qubit is shown in a different color. Thicker black edges indicate connections between program qubits.
	%journal: add colors to couplers connecting problem qubits
}
	\label{fig:contracted_graph_cropped}
\end{figure}

\cref{fig:contracted_graph_cropped} illustrate the structures of the QAC encoding. 
The native embedding that is compatible with both QAC and RBM methods is shown on \cref{fig:PQA_subgraphs}. Each Ising graph has 95 nodes and 125 edges. As we can see, the QAC encoding results in a very sparse graph where less than $10\%$ of the qubits and less than $2\%$ of the couplers are being used.  
\later{Due to the requirements that the problem structure has to be compatible with both QAC and RBM requirements, the Ising graph contains only 95 nodes and 125 edges out of 5627 active qubits and 40279 active couplers of the D-Wave Advantage~4.1 QPU.
Contracted graph (QAC only) has 446 nodes and 1047 edges. Provide a table comparison between hardware graph, QAC, RBM, and QAC+RBM. Can also show pictures of each.}

%In order to run experiments with RBM on larger graphs, we will next drop the requirement of compatibility with the QAC encoding. The resulting problems will be used to compare RBM against QA without error mitigation.

\section{Experiments}\label{sec:experiments}
\subsection{Experimental set-up}\label{sec:exp_setup}
For our experimental analysis, we use the \texttt{Advantage\_system4.1} quantum annealer, which is made available via the Leap quantum cloud service of D-Wave. The annealing process is controlled by the following required annealing parameters: 
$\texttt{num\_reads}=100$, which specifies the number of samples returned per annealer call, and $\texttt{annealing\_time}=100$, which specifies the annealing time in microseconds. All other parameters of the quantum annealer are left at their default settings unless specifically indicated otherwise.
 
For generating test problems for the experiments, we use the following procedure.

(i) Determine the order, $k$, of the error mitigation scheme, which is defined as the number of replicas for RBM and the size of logical qubit for QAC. We use $k=4$ for comparing RBM against QAC and the standard method, and $k\in\{2,4,8\}$, when comparing  RBM against the standard method on large native graphs. 

(ii) Use the method described in the previous section to determine the Ising graph, i.e., the structure of the Ising problem, for the chosen value of $k$.

(iii) For each Ising graph,  generate the coefficients of an Ising problem that has the corresponding structure (non-zero $J_{ij}$ coefficients).
	
In order to generate the coefficients, we can just use  numbers randomly generated in the interval $(0,1)$. But, in such a case, we would not be able to know the optimal solutions for the largest problems since they will be too difficult to solve with exact classic solvers. For that reason, we want to be able to generate problems with planted (known) solutions. Our problem generation algorithm is based on the frustrated loops method \cite{hen2015probing}.
That method constructs an Ising problem on the hardware graph by first generating a random spin configuration as the planted solution. Then, random loops of edges are constructed on the hardware graph, and a frustrated Ising Hamiltonian clause is defined on each loop such that it is minimized by the projection of the planted solution on that loop. One coupling per loop is flipped to introduce frustration, e.g., assign $-\alpha$ to all the couplers in a loop and then flipping the value of one of the couplers to $+\alpha$. (Note that a negative coupler favors adjacent qubits to align their signs, while a positive one encourages them to assume opposite signs.)
The total Ising problem is the sum of all the loop problems, so the planted solution remains a ground state. The hardness of the problems can be tuned by adjusting the clause density, defined as the number of loop clauses divided by the number of spins. For instance, in the case of Chimera-type hardware graph, \cite{hen2015probing} determines that density of about 0.2 results in the hardest problems. 

However, in the case where QAC is used, the communication graph as illustrated on Figure \ref{fig:PQA_subgraphs} is quite sparse, with an average node degree of 2.63, compared to the degrees of approximately 6 and 15 observed for Chimera and Pegasus communication structures, respectively. This reduced degree implies that when loops are generated by random walks, as is the case with the algorithm from \cite{hen2015probing} and its variants, there's a higher likelihood of certain edges being used in multiple loops or remaining absent in any loop. Both scenarios result in relatively easy problem instances. %journal also, some +1 and -1 biases can cancel out

For this reason, we modify the way the random loops are generated. Instead of using random walks, we convert the communication graph into an Eulerian graph by adding a minimum number of parallel edges, and then cover the resulting graph by loops so that each edge/coupler (original or added) is used in exactly one loop. An Eulerian graph, for context, is a graph that can be completely covered by loops where every edge is used exactly once. This revised method generates a loop set where all original graph edges are utilized, with each edge being included in at most two loops.

In order to tune the difficulty levels of our generated problems, we utilize two  parameters concerning the magnitude of coupler values. These parameters are the set of the magnitude values, which sets we call \textit{bias values}, and the probability of picking a member of the set for each particular loop. To get problems of varying difficulties, in this paper we use for bias values the sets $\{9,2\}$, $\{10,2\}$, $\{11,2\}$, and the probability that then larger value is picked is set to $0.08$. For instance, for the $\{9,2\}$ bias values, within any given loop, coupler values are assigned either as $\{-9,\dots, -9,+9\}$ with an $8\%$ probability, where the position of the flipped sign is chosen randomly, or as $\{-2,\dots, -2,+2\}$ with $92\%$ probability.  Finally, for each parameter selection, we generate ten random instances.

\subsection{Comparing RBM with QAC and standard QA}\label{sec:rbm_qac_sqa}
% standard method -- use repetition also (multiple qubits) +++
In this subsection we compare RBM for $k=4$ against the QAC and the standard quantum annealing (SQA) methods on problems generated using the planted solution method described in the previous subsection. Since both RBM and QAC use four physical qubits per logical qubit compared to a single one for SQA, we modify SQA to also use multiple qubits. % as in \cite{Pearson2019}. 
Specifically, 
we use the same four physical qubits to represent a logical one as in the QAC strategy, but apply no penalty bias on the couplers connecting the penalty qubit to the three program ones. After quantum annealing, the value of each logical qubit is determined by a majority vote when the physical qubits' values are different. Unlike QAC, the penalty weight used in each logical qubit is zero, which has the effect that no error mitigation strategy is used. 

However, straightforward implementation of this approach may introduce inaccuracies. In this modified SQA version, the absence of weights on couplers attached to penalty qubits renders each penalty qubit independent, yielding a random value in $\{-1,1\}$ with a $50\%$ probability upon measurement. Consequently, this independence can lead to incorrect outcomes in certain scenarios. Consider a scenario where the three program qubits acquire values $1,1,-1$. Given that the program qubits contain the relevant information, the majority vote should yield a value of $1$ (the majority value) irrespective of the penalty qubit's value. However, since the penalty qubit is formally part of the logical qubit, it might impact the outcome, potentially resulting in an incorrect value of $-1$ with a probability of 0.25. This situation occurs if the penalty qubit obtains a value of $-1$, which happens with a probability of $0.5$, followed by a $0.5$ probability of the majority vote in the set $\{1,1,-1,-1\}$ choosing the incorrect value $-1$ when resolving the tie. 
To resolve this issue, modifying the majority vote function of D-Wave to include only the three program qubits of each logical qubit and ignoring the penalty one suffices.

\begin{figure}
	\centering
	\includegraphics[width=\myfigwidth]{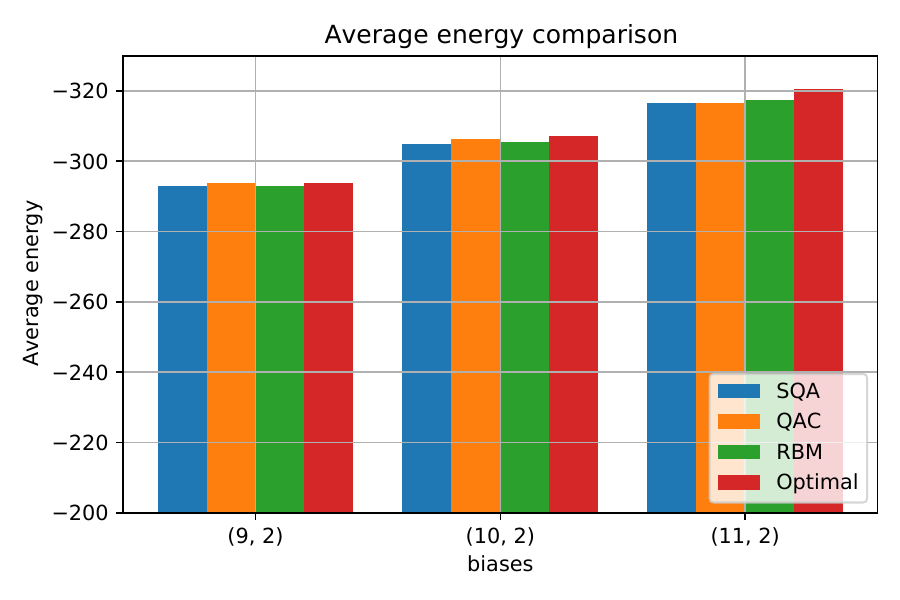}
	\caption{Comparison of the energies of RBM, QAC, and SQA methods.}
	\label{fig:allMethodComparisonPlot}
\end{figure}

The results of the comparison are shown in \cref{fig:allMethodComparisonPlot}. For each of the methods and for each of the three bias parameters, the average of the energies of the methods are shown, as well as the optimal energy corresponding to the planted solution. We observe that in two of the cases QAC performs slightly better than RBM, while RBM performs slightly better than QAC in one of the cases. Only in the easiest of the cases, for a value $(9,2)$ of the bias, one of the methods, QAC, is able to match the energy of the planted solution, while in the other three the average of planted solution energy is strictly lower. SQA performance is the worst of the three methods.

\begin{figure}
	\centering
	\includegraphics[width=\myfigwidth]{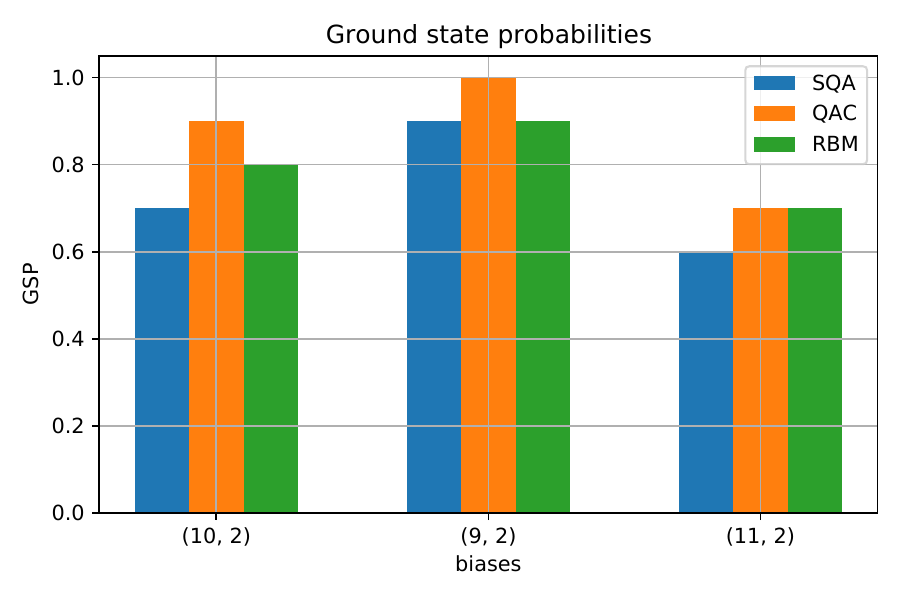}
	\caption{Comparison of the ground state probabilities of RBM, QAC, and SQA methods.}
	\label{fig:allMethodComparisonGSPplot}
\end{figure}

An alternative form of analysis involves comparing the methods not solely based on their attained minimum energy levels, but rather on their respective probabilities of discovering an optimal solution. We define the \textit{ground state probability ({GSP})} of a method $m$ as the ratio between the instances where method $m$ identifies a solution with energy matching the planted solution, and the total number of instances. The results of the experiment are shown on \cref{fig:allMethodComparisonGSPplot}. 
We observe a performance similar to the one found in the previous experiment. 
QAC outperforms RBM in the cases of $(9,2)$ and $(10,2)$ bias values, whereas for the case of $(11,2)$ bias their GSP values are equal. Similarly, RBM outperforms SQA in two cases, while achieving the same GSP in the third scenario.
Our findings align with those reported in \cite{pearson2019analog}, where the methods were evaluated using the earlier generation of D-Wave quantum annealers operating on the Chimera connection architecture instead of the Pegasus platform.

We note that the variations in the performance across the methods in both experiments are rather small. In the first experiment, energy differences within each group remain below $0.4\%$. Similarly, during the second experiment, disparities in the GSPs between methods, barring a single exception, do not exceed $0.1$. This slight difference suggests only a one-count variance in instances where an optimal solution was found.

\subsection{Comparing RBM and standard QA on larger problems}
This section presents a comparison between RBM and the standard quantum annealing (SQA) method. Unlike the input test problems in the previous subsection, these problems do not require compatibility with QAC, allowing the generation of larger and denser problem sets. Specifically, we use values for the number of copies, $k$, within $\{2,4,8\}$, while the average sizes of the Ising problems as well as the sizes of the problems of the previous subsection are given in \cref{tab:1}. %These are the largest problems that could be generated given the methods' requirements and the hardware topology.  
Additionally, our approach requires a modification in the SQA methodology compared to \cref{sec:rbm_qac_sqa}, particularly in how multiple qubits per variable are employed. As we lack $k$ physical qubits to represent a logical one, our method entails employing one of the $k$ identical embedded problems derived from RBM, conducting on it $k$ separate quantum annealings.

\begin{table}[]
\caption{Number of linear and quadratic terms of the Ising problems used in the experiments. In the last three columns, the numbers given are the averages over all clause density (\textit{beta}) values.}
\label{tab:1}
\begin{tabular}{|c||c|c|c|c|}
	\hline
	\# parts     & 4   & 2    & 4    & 8    \\ \hline
	QEC          & yes & no   & no   & no   \\ \hline
	\# linear    & 95  & 2652 & 1219 & 526  \\ \hline
	\# quadratic & 125 & 15349 & 6914 & 2826 \\ \hline
	Figures      & 
	\ref{fig:allMethodComparisonPlot}, \ref{fig:allMethodComparisonGSPplot} &
	\ref{fig:PQA_vs_standard_planted_data_k248Plot}(a), \ref{fig:PQA_vs_standard_planted_data_GSP_k248Plot}(a) &
	\ref{fig:PQA_vs_standard_planted_data_k248Plot}(b), \ref{fig:PQA_vs_standard_planted_data_GSP_k248Plot}(b) &
	\ref{fig:PQA_vs_standard_planted_data_k248Plot}(c), \ref{fig:PQA_vs_standard_planted_data_GSP_k248Plot}(c) \\ \hline 
\end{tabular}
%\bigskip
\end{table}

\begin{figure}
	\centering
	\includegraphics[width=0.6\textwidth]{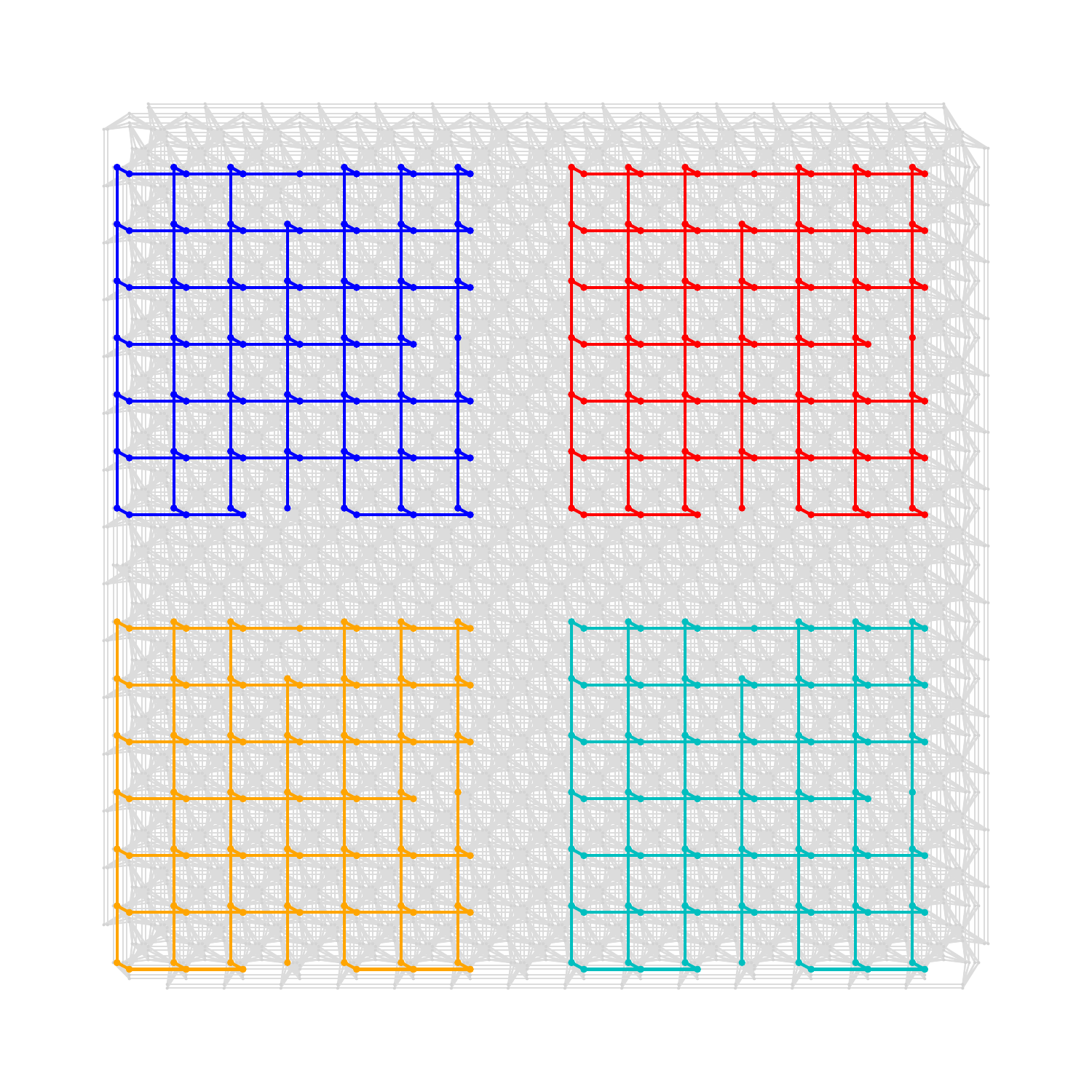}
	\caption{Structure of the Ising graph for the RBM+QAC native embedding for $k=4$. There are $k$ isomorphic Ising graphs embedded in $k$ disjoint areas of the QPU chip. Each of the Ising graphs has structure that allows a QAC native embedding, i.e., to represent each node (logical qubit) as $k$ physical qubits.}
	\label{fig:PQA_subgraphs}
\end{figure}

For each value of $k$, we generate random Ising problems with known optimal solutions using the planted solution method described in the previous section. 
These problems have the same sizes in terms of the numbers of linear and quadratic terms but differ in the value of the coefficients. But because of the much denser Ising graphs underlying the structure of the corresponding Ising problems, e.g., compare \cref{fig:PQA_noQEC_subgraphs} and \cref{fig:PQA_subgraphs}, we will use a different parameter for tuning the hardness of the generated problems. Specifically, we will use a parameter, we call \bet, which determines the clause density of the frustrated loops. We remind that the coefficients of the generated Ising problems are determined as the sum of a set of smaller Ising problems each defined on a loop in the hardware graph. These loops, on the other hand, are determined by a set of randomly generated paths with the property that they cover all the nodes of the hardware graph. The coefficient \bet determines what fraction of the set loops are actually used to define the coefficients of the problem Ising. In Subsection~\ref{sec:exp_setup} above, due to the sparsity of the Ising graph, the hardest problems correspond to $\bet=1$, which is the maximum density of loops, and we had to use other parameters to tune the hardness. However, in the current case of denser Ising problems, using $\bet<1$ may produce harder problems. Hence, in our experiments, we vary the values of \bet in the set $\{0.7,0.8,0.9,1\}$. We also set values $0.08$ and $\{10,2\}$ for the parameters probability and bias values, which were defined in \cref{sec:exp_setup}. For each value of \textit{beta}, we generate $10$ random instances.

\begin{figure}
	\centering
	\includegraphics[width=\myfigwidth]{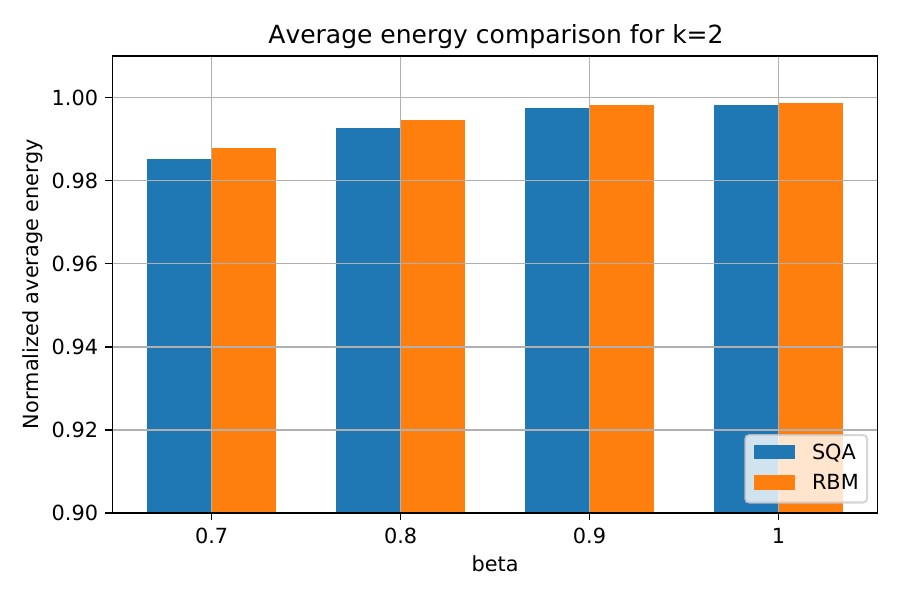}
	\\[-0.2cm]  \hspace*{0.8cm}\small{(a)}	\\ [0.1ex]
	\includegraphics[width=\myfigwidth]{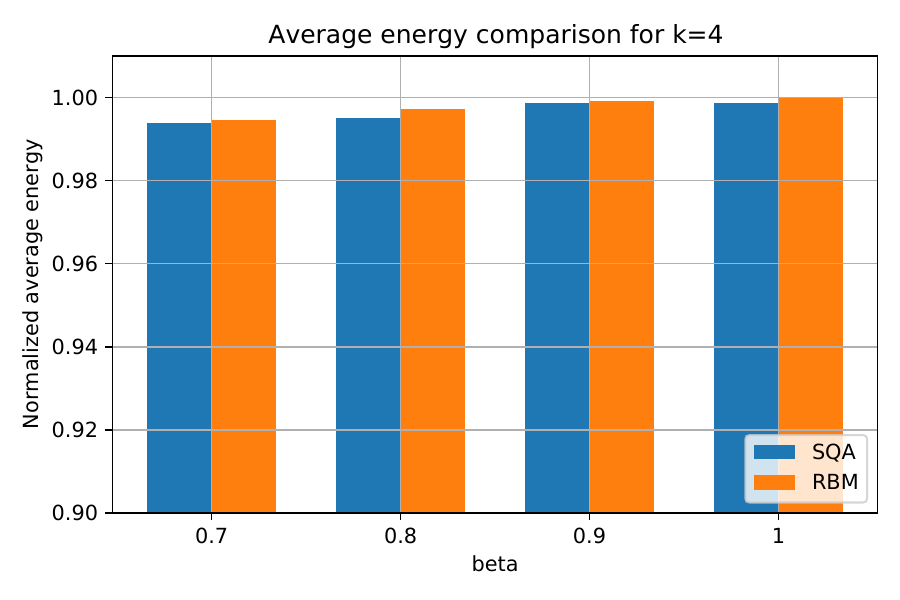}
	\\[-0.2cm]  \hspace*{0.8cm}\small{(b)}		\\ [0.1ex]
	\includegraphics[width=\myfigwidth]{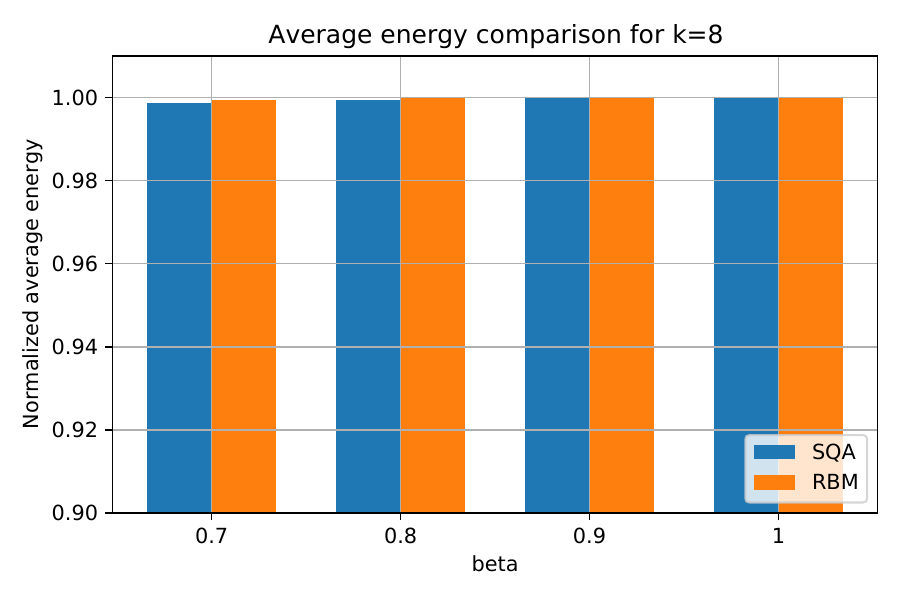}
	\\[-0.2cm]  \hspace*{0.8cm}\small{(c)}	
	\caption{Comparsion of the normalized energies of RBM and SQA methods on Ising problems for $k\in\{2,4,8\}$.}
	\label{fig:PQA_vs_standard_planted_data_k248Plot}
\end{figure}

In the first experiment in this setting, we measure the energy of the best solutions found by the methods. In \cref{fig:PQA_vs_standard_planted_data_k248Plot}, we plot the energies of each of the methods averaged over the set of $10$ Ising problem instances. Since the range of energies vary significantly with $k$ due to the different sizes of the problems, the averaged energies in each group have been normalized by dividing the energy value of the optimal solution. The optimal solution energies have been omitted since they are all equal to one after the normalization.

As one can see on \cref{fig:PQA_vs_standard_planted_data_k248Plot}, in almost all cases RBM outperforms SQA. The two cases where the methods have the same performance with respect to the average energies are for $k=8$ and $\bet\in\{0.9,1\}$, which are the easiest problems. Also, we observe that the accuracies of both algorithms in this experiment get better with increasing $k$. There are two reasons for that. First, by increasing $k$, we increase the number of physical qubits allocated per logical one (by $k$) and thereby the amount of resources used per variable and the accuracy. Second, the size of our Ising problems, and hence their hardness, decreases by $k$. 

\begin{figure}
	\centering
	\includegraphics[width=\myfigwidth]{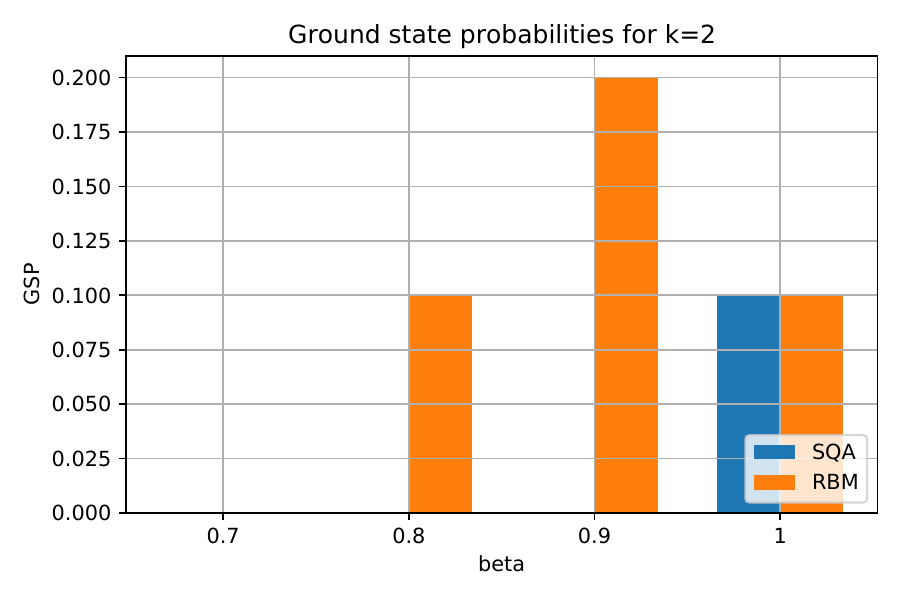}
	\\[-0.2cm]  \hspace*{0.8cm}\small{(a)}	\\ [0.1ex]	\includegraphics[width=\myfigwidth]{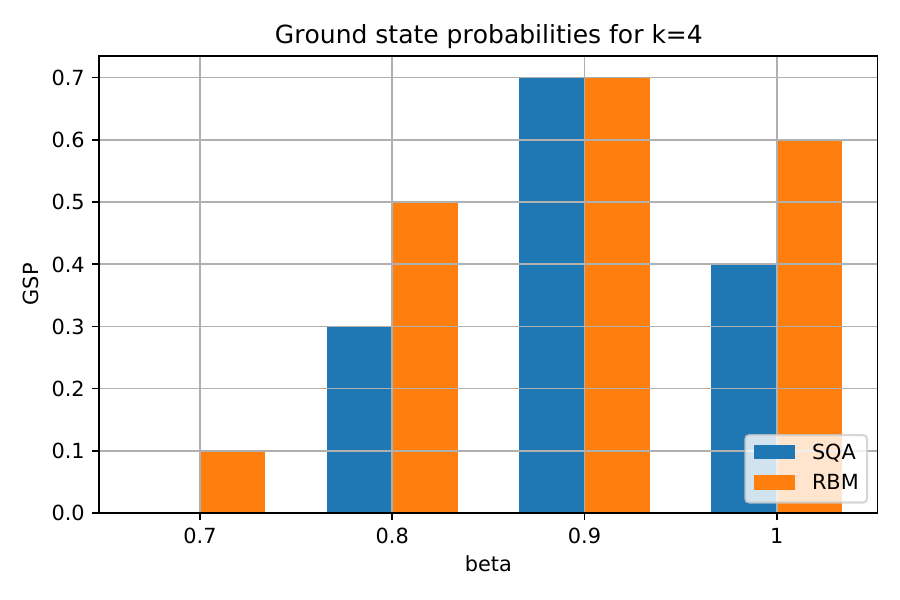}
	\\[-0.2cm]  \hspace*{0.8cm}\small{(b)}	\\ [0.1ex]	\includegraphics[width=\myfigwidth]{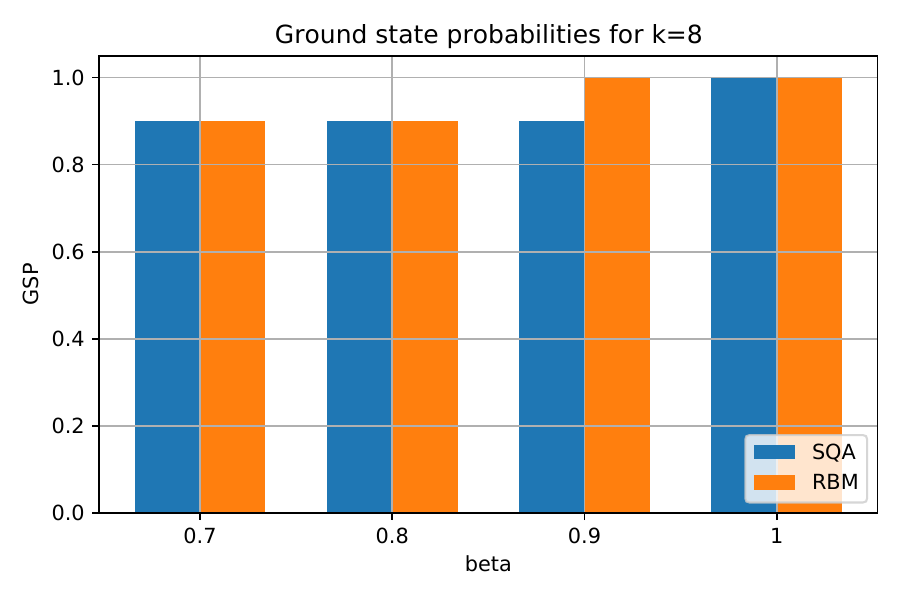}
	\\[-0.2cm]  \hspace*{0.8cm}\small{(c)}	
	\caption{Comparsion of the ground state probabilities of RBM and SQA methods on  Ising problems for $k\in\{2,4,8\}$.}
	\label{fig:PQA_vs_standard_planted_data_GSP_k248Plot}
\end{figure}

An alternative approach to assess the methods is by evaluating their ground state probability (GSP), indicating their likelihood to discover the optimal solution for varying input problem types. The corresponding results are depicted in \cref{fig:PQA_vs_standard_planted_data_GSP_k248Plot}. Notably, this criterion appears to accentuate the increasing difficulty of problems as the parameter $k$ decreases, amplifying the performance disparity between RBM and SQA. For the scenario with $k=8$, RBM and SQA exhibit identical GSP, except at $\bet=0.9$. As for $k=4$, the least challenging problems seem to emerge at $\bet=0.9$, and across the remaining cases, RBM consistently outperforms SQA. A similar trend is observed at $k=2$, wherein RBM successfully identifies the optimal solution for three out of four $\bet$ values, while SQA achieves this in only one case.

We note that RBM and SQA use for the same amount of hardware (qubits and couplers) per variable, but while SQA has the Ising problem embedded just once and performs quantum annealing $k$ times, RBM has the Ising problem embedded $k$ times in the quantum chip and performs quantum annealing just ones. So while in both methods the same logical problem is solved $k$ times, the advantage of RBM comes from the fact that having problems embedded in different parts of the quantum chip helps alleviate any hardware biases.

In our experiments we don't report times as they are virtually the same, as the parameter \texttt{annealing\_time} is set to $100$ for all experiments. The only exception is th implementation of SQA from this section, which has $k$ time larger running time, due to the fact that it involves running the same problem $k$ times on the quantum annealer.

In our experiments, we do not present time measurements as they remain nearly identical across all experiments. This consistency arises from setting the parameter \texttt{annealing\_time} to a constant value of $100$ for each experiment. The sole exception pertains to the implementation of SQA introduced in this section. Notably, SQA involves $k$ times longer runtime owing to the necessity of executing the same problem $k$ times on the quantum annealer.

\section{Conclusions}\label{sec:conclusion}
In this work, we introduced and analyzed a new quantum annealing error mitigation approach called replication-based mitigation (RBM). The key idea behind RBM is to leverage parallel quantum annealing to simultaneously solve multiple copies of the optimization problem embedded across the quantum chip. This helps addresses biases and noise in analog quantum computers.

We have experimentally compared RBM against the quantum annealing correction (QAC) method on problems with structure compatible with both techniques on D-Wave Advantage computers using the Pegasus graph communication structure. The results demonstrate comparable performance between RBM and QAC in terms of solution quality although QAC performs slightly better. However, RBM provides greater flexibility in the connectivity of problems it can handle and can be effectively employed for significantly larger problem sizes.

We have also assessed RBM against standard quantum annealing without error correction on larger problem sizes, which sizes are possible when the structure of the problem does not nee to be compatible with QAC. The experiments indicate systematic improvements in the quality of solutions returned by RBM across different problem settings. This validates RBM's premise of mitigating hardware errors and biases by using replication and parallelism during the quantum annealing process.

Overall, RBM offers a promising approach well-suited for contemporary noisy intermediate-scale quantum annealers. While slightly less accurate in some cases, it presents several practical advantages over QAC and related methods:
(i) RBM has a much simpler implementation, as Ising problems can be embedded into the quantum annealing chip using standard techniques, just across subregions. 
Conversely, QAC's implementation is notably more complex during the embedding phase due to the necessity of mapping each logical qubit to multiple connected physical qubits and ensuring continued connectivity among subsets of problem qubits.
(ii) RBM readily adapts to future quantum annealers with diverse connection topologies, with minimal to no modification needed.
(iii) RBM requires no tuning of parameters, whereas setting the penalty strengths between qubits is critical for good QAC performance.
(iv) Beyond natively encoded problems, RBM applies to non-Ising formulations that require conversion to Ising models.
Almost all practical problems, such as those with constraints or structures incompatible with the hardware topology, necessitate such preprocessing transformations. Mapping such problems to valid Ising formulations introduces penalty weights between qubits. However, applying additional penalty components proves problematic for QAC and related techniques, as they already use penalty weights between the penalty and the problem qubits. The multiple penalty interactions tend to impair solution accuracy. In contrast, RBM embeds subgraphs independently without reconnecting logical qubits, allowing auxiliary variables and constraints to seamlessly integrate within the replications.
One problem for future work is to implement and analyze the performance of RBM on non-native classes of Ising problems.

\section*{Acknowledgments}

\label{sec:acknowledgments}
%Hidden to preserve anonymity.
%\vspace*{0.5cm}
This work was supported by grant number KP-06-DB-11 of the Bulgarian National Science Fund and by the Laboratory Directed Research and Development program of Los Alamos National Laboratory under project 20210114ER. Los Alamos National Laboratory is operated by Triad National Security, LLC, for the National Nuclear Security Administration of U.S. Department of Energy (contract No.\ 89233218CNA000001).

%%OLD
%This work was supported by the U.S. Department of Energy through the Los Alamos National Laboratory. Los Alamos National Laboratory is operated by Triad National Security, LLC, for the National Nuclear Security Administration of U.S. Department of Energy (Contract No.\ 89233218CNA000001). The research presented in this article was supported by the Laboratory Directed Research and Development program of Los Alamos National Laboratory under project number 20220656ER. 
%%%THIS ONE%%This work was supported by grant number KP-06-DB-11 of the Bulgarian National Science Fund and by grant number \ BG05M2OP001-1.001-0003, financed by the Science and Education for Smart Growth Operational Program (2014-2020) and co-financed by the European Union through the European Structural and Investment Funds. 
%This research used resources provided by the Darwin testbed at Los Alamos National Laboratory (LANL) which is funded by the Computational Systems and Software Environments subprogram of LANL's Advanced Simulation and Computing program (NNSA/DOE). 

\bibliographystyle{plain}
\bibliography{quantum, errorCorrection}

\end{document}